\magnification 1100
\tolerance 5000
\def\pmb#1{\setbox0=\hbox{$#1$}%
    \kern-.025em\copy0\kern-\wd0\kern.05em\copy0\kern-\wd0\kern-.025em
    \raise.0433em\box0}

\def\Msol{\,{\rm m_\odot}}

\def\frac#1/#2{\leavevmode\kern.1em
\raise.5ex\hbox{\the\scriptfont0 #1}\kern-.1em
/\kern-.15em\lower.25ex\hbox{\the\scriptfont0 #2}}

\def\lsim{\, \lower2truept\hbox{${< \atop\hbox{\raise4truept\hbox{$\sim$}}}$}\
,}
\def\gsim{\, \lower2truept\hbox{${> \atop\hbox{\raise4truept\hbox{$\sim$}}}$}\
,}

\centerline{\bf {Observability of Early Evolutionary Phases of Galaxies at mm 
Wavelengths}}
\bigskip
Gianfranco De$\,$Zotti$^{1}$, Paola Mazzei$^{1}$, 
Alberto Franceschini$^{2}$, and Luigi Danese$^{3}$

\bigskip
\bigskip
{\sl
$^1$Osservatorio Astronomico, Vicolo dell'Osservatorio 5, 
I-35122 Padova, Italy 
\smallskip
$^2$Dipartimento di Astronomia, Vicolo 
dell'Osservatorio 5, I-35122 Padova, Italy 
\smallskip
$^3$ International School for Advanced Studies, Via Beirut 2--4, 
I-34013 Trieste, Italy}
\bigskip
\bigskip
\noindent
{\bf{Abstract}}

\smallskip
Several lines of evidence and theoretical arguments suggest 
that a large fraction of starlight is absorbed by interstellar 
dust and re-radiated at far-IR wavelengths, particularly during early 
evolutionary phases of early type galaxies, which may even, under some 
circumstances, experience an optically thick phase. Therefore far-IR to 
mm observations are crucial to understand the galaxy evolution.
The strong K-correction makes surveys at mm wavelengths ideally 
suited for studying high-$z$ galaxies. 
The broad redshift range covered by mm surveys 
at sub-mJy flux limits offers a good chance for gaining important information 
also on the geometry of the Universe.
\bigskip
\noindent
{\bf{1. Introduction}}
\smallskip
The IRAS survey has clearly demonstrated the crucial role of dust in shaping
the spectral energy distribution (SED) of galaxies. A direct comparison of
their local luminosity functions in the optical and in the far-IR shows that,
locally, about 30\% of starlight is reprocessed by dust. 

It is very likely that this fraction was higher in the past, when 
the star formation activity was more intense and also early type galaxies 
possessed a plentyful interstellar medium, which might have been metal 
enriched on a very short timescale, the lifetime of the first generation of 
massive stars.

Thus, the observational study of evolution of dust emission in galaxies 
is crucial 
to understand the evolution of galaxies themselves. 
\bigskip

\noindent
{\bf {2. Luminosity Evolution, Dust Absorption and Far-IR/mm Emission} }
\smallskip
A useful scheme for modelling the evolution of galaxies relies 
on the consideration of two extreme patterns (Sandage 1986): on one side, 
disk galaxies, characterized by 
dissipational collapse, with slow gas depletion, i.e. star formation rate 
(SFR) never much higher than today; on the other side, 
spheroidal galaxies thought 
to have used up most of their gas to form stars in a time short compared 
with the collapse time, i.e. with a spectacularly large 
initial SFR. The evolution of galaxies of different Hubble types can then 
be modelled as a suitable combination of the two basic components. 

This scheme is certainly oversimplified in many respects; in particular, it 
does not allow for important facts such as merging, interactions, star 
formation induced by nuclear activity, and so on. Still, it may be useful 
to sketch out some of the chief features. 

Our approach (Mazzei et al. 1992, 1994) exploits chemical and 
photometric evolution models of stellar populations, complemented by 
allowing for the effect of dust. Simple assumptions are adopted for the 
dust component, namely:
the dust to gas ratio is proportional to the metallicity (i.e. 
a constant fraction of metals is locked up in dust grains);
stars and dust are well mixed; 
the ``standard'' grain model (Mathis et al. 1977; Draine and Lee 1984),
including a power law grain size distribution, holds at any time.

If indeed the metallicity and the star formation rate 
in {\it galactic disks} did not vary much throughout their lifetime, 
their SED too remained essentially unchanged, except for optical colours 
being somewhat bluer and the dust being somewhat warmer during the 
early phases (Mazzei et al. 1992).

On the contrary, dramatic far-IR evolution is expected for 
early-type galaxies due to the fast (exponential with a timescale of a few 
Gyr) decrease of the SFR with increasing galactic age. 
Their bolometric luminosity increases by a substantial factor with 
decreasing galactic age, $T$ (a factor $\simeq 10$ from $T = 15\,$Gyr to 
$T = 2\,$Gyr). Moreover, the far-IR to optical luminosity ratio increases 
from local values $\lsim 10^{-2}$ (Mazzei and De Zotti 1994a) 
to $\simeq 1$ or even $\gg 1$ at early times. On the whole, the far-IR 
luminosity of early type galaxies may have increased by about three orders 
of magnitude (if the effect of merging may be ignored), a luminosity 
evolution rate more extreme than even that quoted for optically selected 
quasars.

Under the above assumptions for dust properties, a key parameter in determining
the evolution of the SED of early type galaxies, is the gas consumption rate:
in the case of a fast conversion of gas into stars, the far-IR emission is
never dominant; but if the gas depletion is slower, the galaxy may experience a
prolonged opaque phase, with most of the luminosity emitted in the far-IR
(Mazzei and De Zotti 1996). 

In any case, {\it a substantial dust emission is expected during early 
evolutionary phases of all galaxies.} 
\medskip

\noindent
{\bf{2.1 Are primeval galaxies heavily obscured?} }

\smallskip
As noted above, under some circumstances, galaxies which are in the processes 
of transforming into stars a large fraction of their mass in a relatively 
short time (the conventional definition of primeval galaxies) may be 
heavily obscured by dust. We are unable to work out an a priori estimate 
of how frequently this may occur. 

However, as shown by Franceschini et al. (1994), under the assumption that, 
during the phases of 
intense star formation, most of the optical radiation was absorbed by dust and 
reradiated in the far-IR, a consistent picture obtains in the framework 
of simple luminosity evolution models. In particular, we may account 
for: 
the remarkable lack of high redshift galaxies in optically selected samples 
down to $B\simeq 24$ (Colless et al. 1993; Cowie et al. 1991);
the failure to detect Ly$\alpha$ 
emission in searches for primeval galaxies (e.g., Thompson et al. 1995); 
the deep $60\,\mu$m IRAS counts and, 
exploiting the far-IR/radio correlation for galaxies, most 
of the observed sub-mJy flattening of radio counts over a couple of decades 
in flux.

Also, contrary to recent claims (e.g. Thompson et al. 1995), predictions of 
models entailing strongly obscured primeval galaxies are not in 
conflict, but may even be supported by COBE data on the far-IR to mm 
background (cf. Fig. 1).
\medskip
\noindent
{\bf{2.2  Direct Evidences of Large Amounts of Dust in High-z Sources}}
\smallskip
The observed spectral energy distribution of the ultraluminous object IRAS 
F10214$+$4724 (Rowan-Robinson et al. 1991; Lawrence et al. 1993), 
at $z \simeq 2.3$  is 
remarkably well fitted by a model for young (age $\lsim 1\,$Gyr) spheroidal 
galaxies with strong dust extinction (Mazzei and De Zotti 1994b).

On the other hand, this source hosts an active nucleus which may well be the 
main energy source, as strongly suggested by the evidences of 
gravitational lensing (Eisenhardt et al. 1996). 
Indeed, Granato et al. (1996) obtain a good fit of its SED 
with the dusty torus model that fits the SEDs of 
both broad and narrow line AGNs in the framework of the unified model 
(Granato \& Danese 1994). The diameter of the far-IR emitting dusty torus 
is, in their model, of $\simeq 2\,$kpc (for $H_0 =50$), corresponding 
to an angular size of $0''.3$.

Dust masses $\simeq 10^8$--$10^9\Msol$ (suggesting gas masses of between
$10^{10}$ and $10^{12}\Msol$) are indicated by mm/sub-mm detections (Dunlop et
al. 1994; Chini \& Kr\"ugel 1994; Ivison 1995) of the high $z$ radio galaxies
4C41.17 ($z = 3.8$), 53W002 ($z=2.39$), and 8C1435$+$635 ($z = 4.26$); such
dust masses are 1--2 orders of magnitude higher than found for nearby radio
galaxies (Knapp et al. 1990; Knapp \& Patten 1991). 

Furthermore, evidences of vast reservoirs of dust at high $z$ are 
provided by mm/sub-mm detections of a number of distant radio quiet QSOs 
(Andreani et al. 1993; Ivison 1995, and references therein); several high-$z$ 
radio loud QSOs were also detected, but the 
observed mm fluxes could be accounted for by synchrotron emission.

As in the case of IRAS F10214$+$4724, the relative importance of the nucleus 
and of a possible gigantic burst of star formation in heating the dust 
is still unclear. A better determination of the effective dust temperature, 
that may soon be provided by ISO observations, will certainly be extremely 
helpful: dust temperatures exceeding $\simeq 60\,$K are probably difficult 
to explain with a starbust model. But for a firm assessment of the problem, 
we will probably need the sensitivity and the angular resolution of the 
large mm array which would allow an observational characterization 
of the structure and luminosity of both the torus and the starburst.
\bigskip

\noindent
{\bf{3. Flux vs Redshift at mm Wavelengths}}
\smallskip
The spectral energy distribution (SED) of galaxies during the early 
evolutionary phases, characterized by intense star formation activity, 
is likely to be qualitatively similar to that 
of the nearby starburst galaxy M$\,82$, shown in Figure 2. 

Above a few mm, the luminosity 
is dominated by radio emission. The study of this emission and of its 
evolution is an interesting problem {\it per se} since  
a better understanding of the relative contributions of 
synchrotron and free-free processes at these wavelengths would shed 
light on the properties of the magnetic field on one side and of 
HII regions on the other.

In the range $100\,\mu$m--$1\,$mm, dust emission dominates and 
the continuum spectrum of star forming galaxies  
can be described by a power law of the form $L_\nu \propto \nu^\alpha$ 
with $\alpha \simeq 3.5$ (Franceschini and Andreani 1995; Chini et al. 
1995), although the far-IR to optical luminosity ratio in the case of normal 
late type galaxies is substantially smaller than for M$\,82$, consistent 
with the correlation between far-IR emission and star formation rate. 

Then, as we observe at mm wavelengths galaxies at larger and larger redshifts, 
the K-correction works to rapidly increase the flux, as we move to higher 
and higher frequencies along a steeply increasing spectrum 
(Franceschini et al. 1991; Blain and Longair 1993). 
For steep enough spectral indices, the K-correction 
eventually overcomes the effect of increasing distance, so that, above 
some redshift $z_m$, the flux actually increases with distance.

In the absence of any luminosity evolution, if $\alpha = 3.5$
we have $z_m=2$, 1.83, 1.60, 1.29, 
and 0.96 for $\Omega = 0$, 0.03, 0.1, 0.3, and 1, respectively 
(De Zotti et al. 1996). 

The minimum, however, is shallow, so that it can hardly be 
exploited to determine the density parameter. On the other hand, 
the weak dependence of flux on distance implies a rather uniform distribution 
of sources over a broad redshift range (up to $z \simeq 10$, if galaxies 
were already present) thus offering a good chance of exploring 
the geometry of the universe, particularly once the photometric 
evolution of galaxies will be reasonably well understood.
\bigskip

\noindent
{\bf{4. Predictions for Deep Millimeter Surveys}}
\smallskip

Estimates have been attempted by Franceschini et al. (1991), Blain \& Longair 
(1993, 1996), Mazzei et al. (1996), based on different evolution models. 

Such estimates, however, are a very delicate exercise because they require 
large extrapolations with very limited observational constraints. 
The nearest counts on the far-IR side, the IRAS counts at $60\,\mu$m, span 
a limited range of flux and are rather uncertain at the faint end. On one hand,
the redshift survey by Ashby et al. (1996) of the deep IRAS field at the 
North ecliptic pole (Hacking \& Houck 1987) has discovered that the counts 
may be significantly above average because of 
the presence of a  large supercluster at $z = 0.088$.  At the other hand, 
Gregorich et al. (1995), from a study of a set of deep IRAS fields covering 
a total area about three times larger than that of Hacking and Houck (1987) 
report faint counts about a factor of two higher (note, however, 
that the completeness limit adopted by Gregorich et al. (1995), 
$\simeq 50\,$mJy, is only 2.5 times higher than the estimated rms confusion 
noise; there is thus a serious danger that counts are overestimated at the 
faint end because of source confusion). 

Also, there is a considerable spread in the observed 
$(1.3\,\hbox{mm}/60\,\mu\hbox{m})$ flux ratios of galaxies (cf. Fig. 3), 
implying a correspondingly large uncertainty in the estimated local 
luminosity function of galaxies at mm wavelenghts; this 
uncertainty is strongly amplified by the extreme steepness of the counts.

From a theoretical point of view, there is a great deal of uncertainty on 
the physical processes governing galaxy formation and evolution. At one 
extreme there are models assuming that the comoving density of galaxies 
remained essentially constant after their formation, while they evolved 
in luminosity due to the ageing of stellar populations and the birth 
of new generations of stars (pure luminosity evolution). At the other extreme, 
according to some hierarchical clustering models, large galaxies are formed by 
coalescence of large numbers of smaller objects. The observed properties 
of both disk and spheroidal galaxies imply that extensive merging cannot 
have occurred in the last several billion years (cf. e.g. Franceschini et al. 
1994); it must, therefore, have occurred at significant redshifts.

The faint end of expected counts may be strongly different in the two cases. 
Of course, models advocating extensive merging at $z \ge 1$--2 imply a 
sharp enhancement of the counts at faint flux densities (Blain and Longair 
1993, 1996). The flux density at which such enhancement begins is model 
dependent and is sensitive to several unknown quantities such as the 
merging history, the star formation rate and the initial mass function 
during the merging process, the dust temperature. On the other hand, 
a sub-$L_\star$ galaxy with a dust temperature 
of $60\,$K, typical of galaxies with very intense star formation, at 
$z\ge 1$ has a flux density $< 10\,\mu$Jy at $1.3\,$mm. 

All in all, the actual physical processes governing the formation and the 
early evolution of galaxies may be very complex and may depend on an 
impressive number of unknown or poorly known parameters: spectrum of 
primordial density perturbations, hydrodynamic processes in the primordial gas, 
merging rate, star formation rate, 
initial mass function, galactic winds, infall, interactions, dust 
properties, and so on. The observational contraints are still very poor. Hence, 
a direct observational study of these phases is essential. 

If, as argued 
above, metal enrichment and condensation of metals into dust grains 
occurs very quickly in primeval galaxies, far-IR to mm observations 
will play a crucial role in this field. ISO and SCUBA offer excellent 
prospects; their data will certainly help very much to discriminate 
between different scenarios. On the other hand, the much better sensitivity 
of the large mm array is required to test the possibility of substantial 
merging at $z > 1$--2, as expected, in particular, in the framework of 
cold dark matter cosmologies (Kauffmann et al. 1993). 
\bigskip
\noindent
{\bf{5. Conclusions}}
\smallskip
The emission from interstellar dust, which locally comprises $\simeq 30\%$ of
the global bolometric luminosity of galaxies, is very likely to have   been
significantly larger during earlier evolutionary phases, when the   (metal
enriched) ISM was more abundant. Therefore observations of the   dust emission
are crucial to understand the galaxy evolution. 

In the case of early type galaxies, very poor of dust and gas at present,  the
evolution in the far-IR/mm bands could have been very spectacular,   to the
point that during the first 1--2 Gyr of their lifetime, most of   their
starlight could have been reprocessed by dust; this corresponds   to an
increase of the far-IR luminosity by more than three orders of magnitude. 

A related issue is the primary power source of ultraluminous IRAS sources (and
in particular IRAS F$10214+4724$) and of the very large far-IR/mm emission from
some high-z radiogalaxies. From the far-IR luminosity and the available
constraints on dust temperature it is concluded that the dust distribution in
the most luminous sources has a size of at least several hundred parsecs
($H_0=50$). The planned large mm array might resolve these sources even if  
the far-IR emission comes from a dusty torus surrounding an active nucleus. 

The very steep increase of spectra of galaxies with increasing frequency   in
the range few-mm to $\simeq 100\,\mu$m makes surveys at mm wavelengths  
exceptionally well suited for investigating the evolution of galaxies up to
$z\simeq 10$. The very weak dependence of flux on distance for $z>1$ implies
a very uniform coverage of the full redshift range over which   galaxies
presumably exist; once the physics of the evolution is understood, this gives
a good chance of investigating also the geometry of the   universe. 
  
It may be debated whether the large mm array is the appropriate instrument for
surveys aimed at investigating the early evolutionary phases of galaxies.
Indeed SCUBA is expected to be capable of following the evolution of dust
emission from bright galaxies up to very high redshifts. On the other hand,
substantially better sensitivity and spatial resolution than achievable with
SCUBA may be necessary to investigate e.g. the process of extensive merging of
small dusty clumps at $z\gsim 1$--2. Other methods, such as the study of
absorption lines in the spectra of high-z quasars, may be heavily biased if the
clumps are very dusty. 

On the other hand, for such surveys a not too small field of view at 
$\simeq 1\,$mm is essential.

\medskip\noindent
Work supported in part by ASI.
\bigskip
\noindent
{\bf{Referencens}}
\bigskip

Andreani, P., La Franca, F., Cristiani, S. (1993): MNRAS {\bf 261}, L35

Ashby, M.L.N., Hacking, P.B., Houck, J.R., Soifer, B.T., Weisstein,
E.W.  (1996): ApJ {\bf 456}, 428

Blain, A.W., Longair, M.S. (1993): MNRAS {\bf 264}, 509

Blain, A.W., Longair, M.S. (1996): MNRAS in press

Chini, R., Kr\"ugel, E. (1994): A\&A {\bf 288}, L33

Chini, R., Kr\"ugel, E., Lemke, R., Ward-Thompson, D. (1995): A\&A 
{\bf 295}, 317

Cohen, M., Volk, K. (1989): AJ {\bf 98}, 1563 

Colless, M., Ellis, R., Taylor, K., Hook, R. (1993): MNRAS {\bf 261}, 
19

Cowie, L.L., Songaila, A., Hu, E.M. (1991): Nat {\bf 354}, 460 

De Zotti, G., Franceschini, A., Mazzei, P., Toffolatti, L., Danese, 
L. (1996): Ap. Lett. \& Comm. in press

Draine, B.T., Lee, H.M. (1984): ApJ {\bf 285}, 89

Dunlop, J.S., Hughes, D.H., Rawlings, S., Eales, S.A., Ward, M.J. 
(1994): Nat {\bf 370}, 347

Eisenhardt, P., Soifer, B.T., Armus, L., Hogg, D., Neugebauer, G., 
Werner, M.: (1996), ApJ, in press

Franceschini, A., Andreani, P. (1995): ApJ {\bf 440}, L5

Franceschini, A., Granato, G.L., Mazzei, P., Danese, L., 
De Zotti, G. (1995): Proc. COBE Workshop on ``Unveiling the Cosmic IR 
Background'', College Park, MD

Franceschini, A., Mazzei, P., De Zotti, G., Danese, L. (1994): 
ApJ {\bf 427}, 140

Granato, G.L., Danese, L. (1994): MNRAS {\bf 268}, 235

Granato, G.L., Danese, L., Franceschini, A. (1996): ApJ in press

Gregorich, D.T., Neugebauer, G., Soifer, B.T., Gunn, J.E., 
Herter, T.L. (1995): AJ {\bf 110}, 259

Hacking, P., Houck, J.R. (1987): ApJS {\bf 63}, 311

Hauser, M. (1995): Proc. COBE Workshop on ``Unveiling the Cosmic IR 
Background'', College Park, MD

Huang, Z.P., Thuan, T.X., Chevalier, R.A., Condon, J.J., Yin, Q.F. 
(1994): ApJ {\bf 424}, 114

Hughes, D.H., Gear, W.K., Robson, E.I. (1989): MNRAS {\bf 244}, 759 

Kauffmann, G., White, S.D.M., Guiderdoni, B. (1993): MNRAS {\bf 264}, 
201

Ivison, R.J. (1995): MNRAS {\bf 275}, L33

Kennicutt, R.C. Jr. (1992): ApJ {\bf 388}, 310  

Knapp, G.R., Bies, W.E., van Gork, J.H. (1990): AJ {\bf 99}, 476

Knapp, G.R., Patten, B.M. (1991): AJ {\bf 101}, 1609

Lawrence, A., et al. (1993): MNRAS {\bf 260}, 28

Mathis, J.S., Rumpl, W., Nordsieck, K.H. (1977): ApJ {\bf 217}, 425

Mazzei, P., De Zotti, G. (1994a): MNRAS {\bf 266}, L5

Mazzei, P., De Zotti, G. (1994b): ApJ {\bf 426}, 97

Mazzei, P., De Zotti, G. (1996): MNRAS, in press

Mazzei, P., De Zotti, G., Xu, C. (1994): ApJ {\bf 422}, 81

Mazzei, P., Lonsdale, C., Chokshi, A. (1996): in preparation

Mazzei, P., Xu, C., De Zotti,\ G. (1992): ApJ {\bf 256}, 45

Puget, J.L., Abergel, A., Bernard, J.-P., Boulanger,F., Burton, 
W.B., D\'esert, F.-X., Hartmann, D. (1996): A\&A, in press

Rowan-Robinson, M., et al. (1991): Nat {\bf 351}, 719 

Sandage, A. (1986): A\&A {\bf 161}, 89  

Thompson, D., Djorgovski, S.G. (1995): AJ {\bf 110}, 982

Thompson, D., Djorgovski, S., Trauger, J. (1995): AJ {\bf 110}, 963

\vfill\eject

\noindent
{\bf {Captions}}
\medskip
\noindent
Fig. 1: contributions to the far-IR to mm background intensity predicted by two
different models for galaxy evolution, compared with limits (Hauser 1995) and
estimates (dotted lines, Puget et al. 1996)  based on COBE data. The lower thin
curve is a minimal contribution from normal galaxies, estimated assuming no
evolution up to $z=1$. The dashed line is a {\it merging} model, in which most
galaxies form stars and build up at $z \simeq 1$. The thick continuous line is
an evolution model at constant galaxy mass function (for more details on both
models see Franceschini et al. 1995). 
\medskip
\noindent
Fig. 2: observed continuum spectrum of M$\,82$ from UV to radio wavelengths. 
Data are from Hughes et al. (1989) and references therein, Cohen and Volk 
(1989), Huang et al (1994) and references therein, Kennicut (1992).

\medskip
\noindent
Fig. 3: SEDs of galaxies observed at mm wavelengths by Chini et al. (1995), 
normalized to $60\,\mu$m fluxes.

\vfill\eject
\bye